\documentclass[10pt,prd,nofootinbib,reprint,superscriptaddress,longbibliography]{revtex4-2}

\usepackage{amsmath, amssymb, amsthm, graphicx, epsfig, fancyhdr,epsfig}

\usepackage{tikz-feynman}
\tikzfeynmanset{compat=1.1.0}

\usepackage{tikzsymbols}
\usepackage{natbib}
\usepackage{xcolor}
\usepackage{ulem}

\makeatletter
\def\alt{\mathrel{\mathpalette\gl@align<}}
\def\agt{\mathrel{\mathpalette\gl@align>}}
\def\gl@align#1#2{\lower.6ex\vbox{\baselineskip\z@skip\lineskip\z@
\ialign{$\m@th#1\hfil##\hfil$\crcr#2\crcr\sim\crcr}}} \makeatother

\def\beq{\begin{equation}}
\def\eeq{\end{equation}}
\def\bea{\begin{eqnarray}}
\def\eea{\end{eqnarray}}

\begin{document}

% page numbers bottom-center
\pagestyle{plain}

\title{Constraining Light Dark Photons from GW190517 and GW190426\_152155}

\author{Diptimoy Ghosh}
\email{diptimoy.ghosh@iiserpune.ac.in}

\author{Divya Sachdeva}
\email{divya.sachdeva@students.iiserpune.ac.in}

\affiliation{Department of Physics, Indian Institute of Science Education and Research Pune, India}

%%%%%%%%%%%%%%%%%%%%
%%%%%%%%%%%%%%%%%%%%
\begin{abstract}
Ultralight dark photons predicted in several Standard Model extensions can trigger the superradiant instability around rotating black holes if their Compton wavelength is comparable to the Blackhole radius. Consequently, the angular momentum of the black hole is reduced to a value which depends upon the mass and spin of the black hole as well as the mass of the dark photon. We use the mass and spin measurements of the primary black holes in two recently observed binary black hole systems: GW190517 and $\rm GW190426\_152155$
to constrain dark photon mass in the ranges $1.7\times 10^{-14}{\rm\ eV}<m_{A'}<7.6\times 10^{-13}{\rm\ eV}$ and 
$1.3\times 10^{-13}{\rm\ eV}<m_{A'}<4.2\times 10^{-12}{\rm\ eV}$ respectively, assuming a timescale of a few million years from the time of formation of the binary black hole system to the time of their merger. We also discuss an interesting X-ray binary system, MAXI J1820$\_$070, albeit with a relatively small value of the spin parameter. 
\end{abstract}
\maketitle
%%%%%%%%%%%%%%%%%%%%%%%%%%%%%%%%%%%%%%%%%%%%%
%%%%%%%%%%%%%%%%%%%%%%%%%%%%%%%%%%%%%%%%%%%%%

\section{Introduction}
The presence of dark matter (DM) in the universe and its dominance over luminous matter is well-established. In the absence of any DM candidate within the Standard Model (SM), a variety of models have been proposed to explain this additional matter content by introducing new degrees of freedom with masses ranging all the way down to $10^{-22}$ eV (lower bound comes from studies of dwarf galaxy morphology~\cite{Chen:2016unw}). One promising DM candidate in the ultralight regime is the dark photon (DP), gauge boson of a $U(1)$ gauge group which can acquire its mass through the Higgs or St\"{u}ckelberg mechanism, proposed in various beyond the SM scenarios~\cite{Goodsell:2009xc,Camara:2011jg,Kaneta:2016wvf}. By ultralight DM, we mean particle of masses smaller than $10^{-10}$eV; the mass range $10^{-20}-10^{-22}$eV is often referred to as fuzzy DM~\cite{Hu:2000ke,Hui:2016ltb} and the rest of the range with weak coupling to SM particle as weakly interacting slim particles~\cite{Arias:2012az,Nelson:2011sf,Nakayama:2019rhg}. While these states conform to certain observational features of DM~\cite{Marsh:2013ywa}, they are also relevant in models explaining inflation and dark energy~\cite{Boehmer:2007qa,Koivisto:2008xf}.

Being ultralight, they must be produced non-thermally in the early Universe, for example via the misalignment mechanism~\cite{Nelson:2011sf,AlonsoAlvarez:2019cgw}, parametric resonance production or tachyonic instability of a (pseudo)scalar field~\cite{Agrawal:2018vin,Dror:2018pdh,Co:2018lka}, or from the decay of a cosmic strings~\cite{Long:2019lwl}, depending on how they couple to the SM. The strongest constraints on ultra light gauge bosons with very weak coupling to the SM particles come from equivalence principle tests, including those from the E\"{o}t-Wash group~\cite{PhysRevD.50.3614,PhysRevLett.100.041101} and Lunar Laser Ranging experiments~\cite{PhysRevLett.93.261101}. These are fifth-force experiments that do not assume dark photon  as DM. Along with these, the parameter space  corresponding to larger coupling with SM particles is constrained by a variety of laboratory-based experiments, astrophysical and cosmological observations (see Ref~\cite{Filippi:2020kii} for detailed review).

Despite various efforts, the existence of DM is inferred only through its gravitational effects on visible matter. In this respect, the process of superradiance~\cite{1969NCimR...1..252P,PhysRevD.22.2323,PhysRevD.7.949,PhysRevD.58.064014,PhysRevLett.28.994,1973ApJ...185..649P,Press:1972zz,ZOUROS1979139,Arvanitaki:2009fg,Brito_2015,Brito:2015oca} around Kerr black holes (BHs) provide a unique way of probing ultralight bosons solely via their gravitational interactions. In this phenomenon, a gravitational bound state of the bosonic field develops which leads to a large extraction of energy and angular momentum from a rotating BH if the black hole radius is comparable to the boson's Compton wavelength. Thus, the determination of spin and mass of BH can place a constraint on the existence of the ultralight bosons with a particular mass. There exist several studies that exploit properties of stellar-mass astrophysical BHs determined via their accretion observations or Blackhole binary (BBH) system observed via X-ray or Gravitational wave signatures to unravel the existence of ultralight vector fields. For example, Ref.~\cite{Pani:2012vp,Davoudiasl:2019nlo} exploit supermassive blackholes to disfavour the fuzzy DM mass region, Ref.~\cite{Baryakhtar:2017ngi,Cardoso_2018,Stott:2020gjj} use observations of stellar mass BHs, X-ray binaries, and gravitational wave events to probe light scalar and vector bosons masses of different range. In the present work, we use the mass and spin measurements  of the two recently observed BBH systems viz. GW190517~\cite{Ng:2020ruv} and GW190426 152155~\cite{Li:2020pzv}, and of the X-ray binary MAXI J1820+070~\cite{Zhao:2020cjx} to constrain the parameter space of DP complementing and extending earlier studies. For the case of the first two  BBH systems, spin and mass are obtained from the data released by LIGO-Virgo in their catalog GWTC-2~\cite{Abbott:2020niy} and for the third candidate, analysis is done by fitting the spectra obtained from {\it Insight}-HXMT\footnote{http://hxmten.ihep.ac.cn/}~\cite{2020SCPMA..6349503L,2020SCPMA..6349502Z,2020SCPMA..6349505C,2019arXiv191004451C}.

% ----- Has to be modified -----
 \section{Superradiance Overview} 
We begin by recapitulating the key results related to the process of superradiance for light vector fields that are useful to our analysis. For an in-depth discussion of the subject, numerous resources~\cite{Arvanitaki:2009fg,Pani:2012vp,Witek:2012tr,Pani:2012bp,Brito_2015,Brito:2015oca,PhysRevD.96.024004,East:2017ovw,Baryakhtar:2017ngi} exist.

Light non-relativistic dark photon of mass $m_{A'}$ around a Kerr blackhole of mass $M_{\rm BH}$  with the angular velocity of the BH horizon as $\Omega_H$ may experience a superradiant instability if the corresponding wave mode of frequency $\omega$ satisfies the  following condition~\cite{PhysRevD.7.949,PhysRevD.58.064014,Arvanitaki:2009fg}: 
 \beq
\frac{\omega}{m} < \Omega_H\,,
\label{SRcond}
\eeq
with $m$ being the total angular momentum of boson along the BH's rotating axis and $\Omega_H$ being related to the dimensionless spin parameter $a^*\equiv J_{\rm BH}/(G_NM_{\rm BH}^2)\in[0,1)$ by 
\beq
\Omega_H = \frac{1}{2 G_N M_{\rm BH}}\frac{a^*}{1 + \sqrt{1 - a^{*2}}}\,,
\label{OmegaH}
\eeq
where $G_N$ is Newton's gravitational constant and $J_{BH}$ is the total angular momentum of the BH. For a significantly rotating blackhole, this condition implies that superradiance is effective only if the Compton wavelength of the light vector is comparable to the radius of the  black hole which is given as,
\beq
R_{\rm BH}\,=\,G_NM_{\rm BH}(1\pm \sqrt{1-a_*^2}).
\eeq
For growth of superradiant states, the boson need not have any initial number density; quantum fluctuations are sufficient to populate a boson cloud around the BH. The boson population will grow exponentially with time, extracting energy and angular momentum from the BH. 

\begin{figure}
\centering
\includegraphics[width=0.5\textwidth]{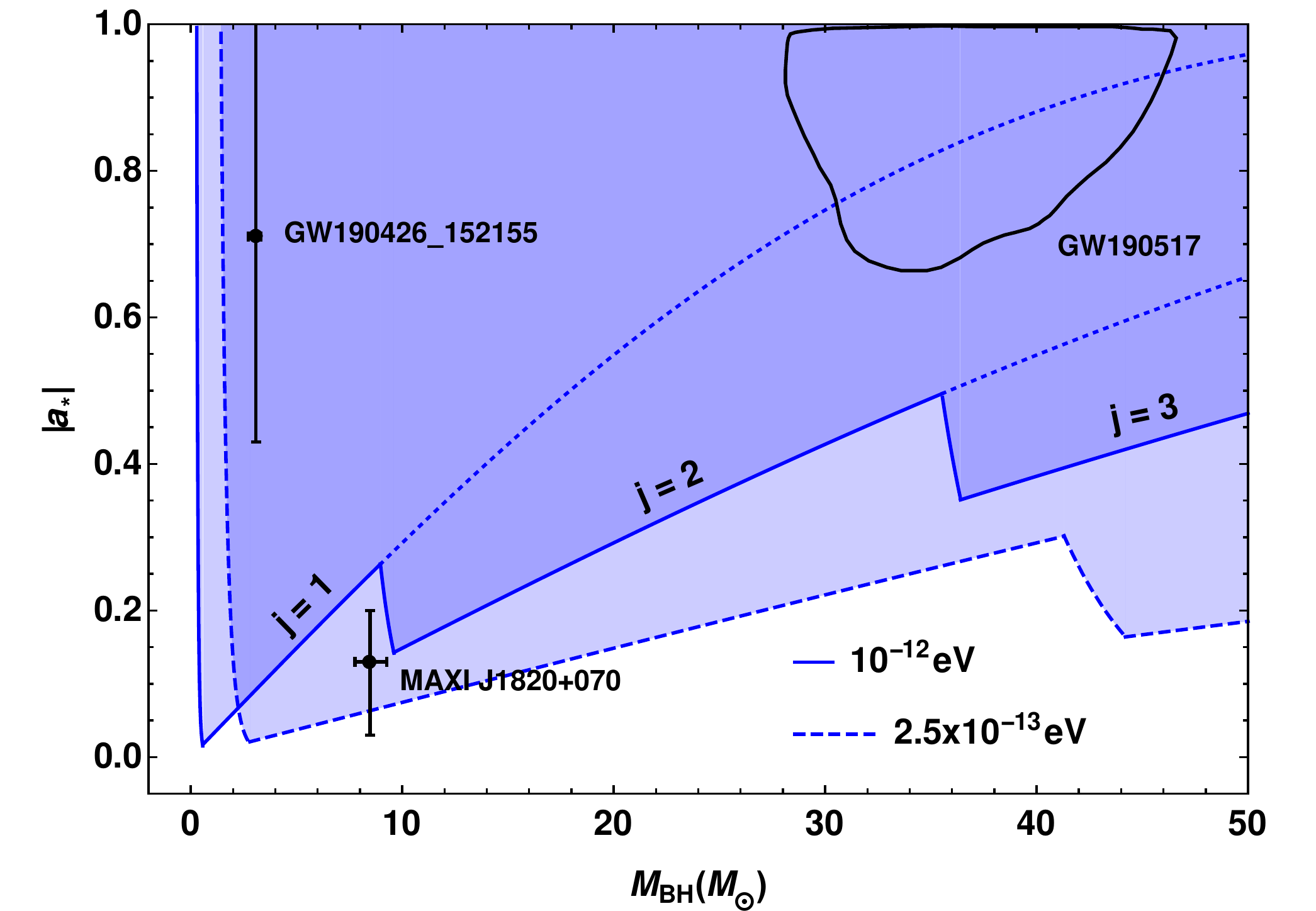}
\caption{The solid and the dashed curves satisfy eqn.~\ref{eq:muV limit} and eqn.~\ref{eq:mu limit} (with an equality sign) for j=1,2,3 level for two representative values of the  vector field mass, $m_{A'}\sim 10^{-12}{\rm eV}$ and $m_{A'}\sim 2.5\times10^{-13}{\rm eV}$ respectively. Each shaded region, thus, satisfies the superradiance condition. The black curve corresponds to $1\sigma$ contour for GW190517 taken from Ref.~\cite{Ng:2020ruv}. The other two data points are the primary BH's mass and spin measurements of the BBH system
$\rm GW190426\_152155$~\cite{Li:2020pzv} and X-ray binary system MAXI J1820+070~\cite{Zhao:2020cjx} with $1\sigma$ errors. A BH excludes the vector boson mass if it lies in the shaded region, within experimental error. }
\label{fig:1d}
\end{figure}

The leading contribution to the growth rate comes from $l=0$ and $j=m=1$ mode~\cite{Arvanitaki:2009fg,Baryakhtar:2017ngi} and is given by 
\begin{align}
\Gamma_V&=4a^*G_N^6M_{\rm BH}^6m_{A'}^7\,.
\label{rate}
\end{align}
For a BH starting with spin $a_{*,0}$ and the cloud extracting its angular momentum by an amount $\Delta a_*\,=\,a_{*,0}-a_*$, the final occupation number for the mode with azimuthal quantum number, $m$, is given by 
\begin{equation}
N_m\simeq\frac{G_NM_{\rm BH}^2\Delta a^*}{m}\, .
\end{equation}
Thus, if we require the superradiance rate to be fast enough, depleting the spin of a BH by $\Delta a_*$ amount within the relevant timescale ($\tau_{\rm BH}$), the following additional condition should be satisfied for each mode:
\begin{equation}
\Gamma_V\,\tau_{\rm BH}\ge\ln N_m\,.
\label{second}
\end{equation} 
Although the $l=0, \, j=m=1$ mode is the dominant one, higher-$m$ modes would be relevant if growth of the $(m-1)^{\rm th}$ level stops within the lifetime of the BH. While for the dominant mode,  analytic expression for the growth rate is available (eqn.~\ref{rate}), this has to be calculated numerically for subdominant modes. Using Ref.~\cite{Baryakhtar:2017ngi}, we calculate the growth rates for different modes of weakly-coupled vector bosons of different masses and illustrate their relevance as a function of the BH mass and spin in fig~\ref{fig:1d}.   

Using eqn.~\ref{second}, an upper limit on the DP mass for an observation of a BH mass and spin can be placed by demanding that superradiance has not depleted
the spin of the BH by $\Delta a^*$,
\begin{align}
m_{A'}&<\left(\frac{\ln N_m}{4a^*G_N^6M_{\rm BH}^6\tau_{\rm BH}}\right)^{1/7}\, . \label{eq:muV limit}
\end{align}
A lower limit can be obtained from eqn.~\ref{SRcond} by requiring that superardiance is not effective enough to reduce the spin:
\begin{equation}
m_{A'}>\Omega_H\,.
\label{eq:mu limit}
\end{equation}
We will see in the next section how observation of BBHs in gravitational wave detectors and in X-rays can constrain specific ranges of dark photon mass using eqn.~\ref{eq:muV limit} and \ref{eq:mu limit}.

 \section{Results}
In this section, we consider the observation of two BBHs systems seen via gravitational wave signature and an X-ray binary as mentioned in the previous section to probe a light DP mass.
Taking a cue from Ref.~\cite{Ng:2019jsx,Ng:2020ruv}, we assume an inspiral timescale of $10^7$~yr from the time the binary black hole system is formed to the time the black holes merge.
%as suggested by simulation studies~\cite{PortegiesZwart:1999nm,10.1093/mnras/%stu824,Morscher_2015,Bavera:2019fkg}.
%Thus, we conservatively take $\tau_{\rm BH}=10^7$ years as our fiducial value. 

For BHs with an astrophysical origin, Thorne has given a upper limit to the reduced spin $a_{*,0}\sim 0.998$~\cite{1974ApJ...191..507T}. This limit comes from the accretion of the surrounding gas on a BH, and its balance with superradiance effects. We take this as the maximum spin a BH can have. The other most important parameters in the present context are the mass and the final spin of the primary BH, i.e the value just before it is  merged. We use the results of the detailed analysis performed in Ref.~\cite{Ng:2020ruv} and Ref.~\cite{Li:2020pzv} which utilise the data provided on BBH systems GW190517 and $\rm GW190426\_152155$ by GWTC-2 to obtain the spin of the primary BH as a function of its mass. These measurements are shown in fig.~\ref{fig:1d}. 
%We depict these in fig.~\ref{fig:1d}. 
%Note that the later candidate has large false alarm rate. 
We use the lower edge of the $1\,\sigma$ contour of GW190517 and $\rm GW190426\_152155$ (as shown in fig.~\ref{fig:1d}) to obtain conservative bounds.

\begin{figure}[t]
\centering
\includegraphics[width=0.47\textwidth]{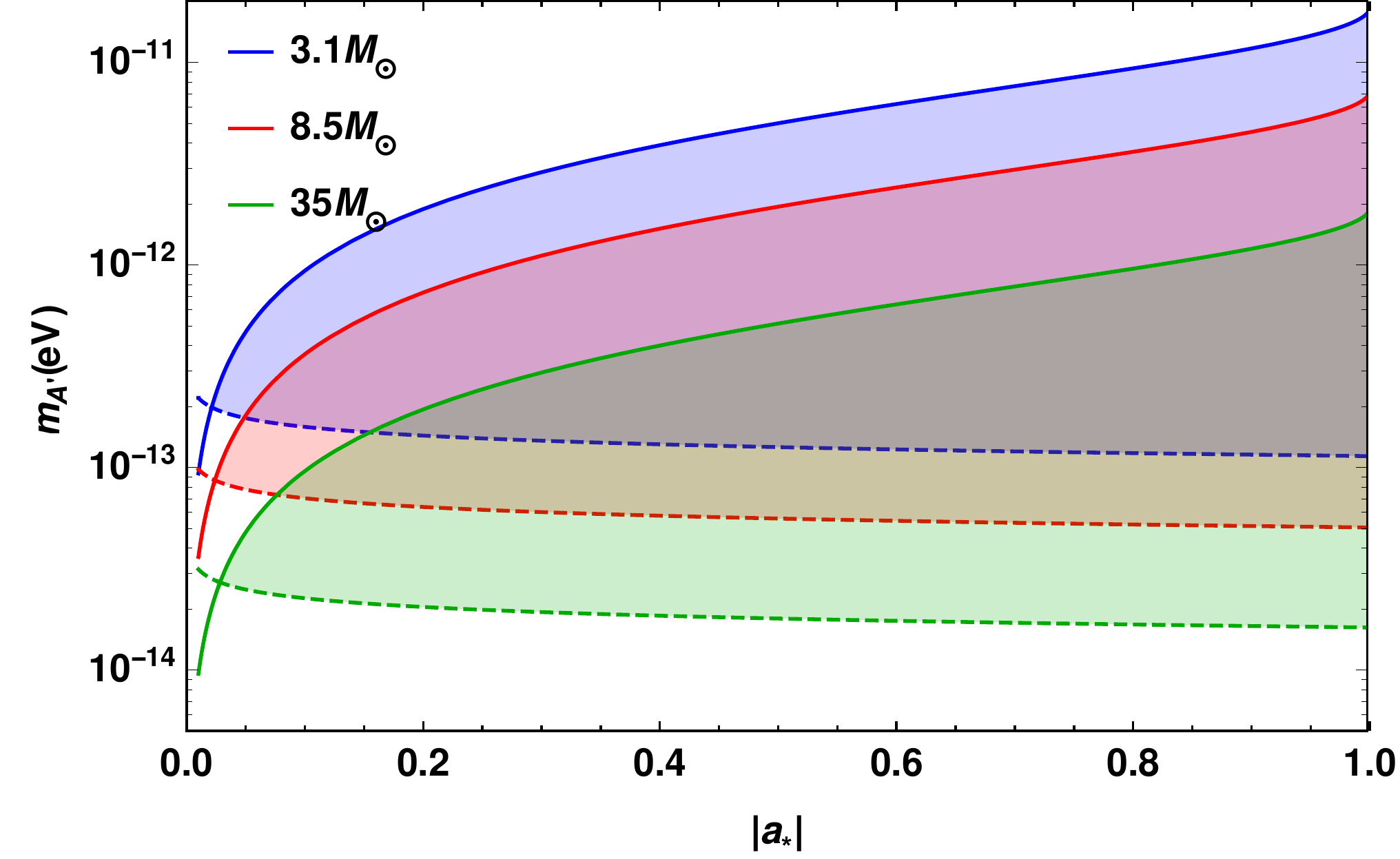}
\caption{Variation of bounds on the DP mass as a function of spin of the primary BH of a BBH system GW190517 (green), $\rm GW190426\_152155$ (blue), and X-ray binary system MAXI  J1820+070 (red) corresponding to the inspiral timescale of $10^7$ yrs. The solid and dashed contours correspond to eqn.~\ref{eq:mu limit} and eqn.~\ref{eq:muV limit} (with an equality sign) respectively. The region between these curves is disfavoured.}
\label{fig:a dependence}
\end{figure}

Following the discussion above and in the previous section, a specific mass range of light vector boson is constrained. In fig.~\ref{fig:1d}, we show the effect of superradiance for representative values of vector boson mass. The curves satisfy eqn.~\ref{eq:muV limit} and eqn.~\ref{eq:mu limit} (with an equality sign) for j=1,2,3 level. As mentioned earlier, for a vector boson, the $j=m=1, \ell=0$ level dominates and thus the bound can be obtained by considering just this level. For GW190517, for some part of the 1$\sigma$ contour (and for larger dark photon mass, see fig.~\ref{fig:1d}), higher-$j$ modes can also contribute. We found numerically that including these increases the upper limit at most by a factor of 2. Below we quote the limits using the $j=1$ level only for which analytic expressions are available.

%For system MAXI J1820+0740, the spin value is small, and considering large uncertainty

We find that the observations of two BBH systems disfavours light bosons in the following ranges,
 \begin{align}
 1.7\times 10^{-14}{\rm\ eV}&<m_{A'}<7.6\times 10^{-13}{\rm\ eV}\,,\\
 1.3\times 10^{-13}{\rm\ eV}&<m_{A'}<4.2\times 10^{-12}{\rm\ eV}\,.
 \end{align}
In fig.~\ref{fig:1d}, we also show the data point for the X-ray binary MAXI J1820\_070. Similar to the case of the other two BBH candidates, using the lower edge (of data point) for spin measurement, we are able to constrain only a small range of DP mass: $8.3\times 10^{-14}{\rm\ eV}<m_{A'}<1.1\times 10^{-13}{\rm\ eV}$.
With better spin measurements in the future, one can improve this bound. It is interesting to note that even observations of relatively lower spin BHs, in principle, can provide useful limits, provided the spin measurement is precise enough.

 To better understand the results and implications of superradiance, fig~\ref{fig:a dependence} shows the variation of the constraints as a function of BH spin for different BH mass. Clearly, if the spin is higher, the constraints on the DP mass will become stronger. With increasing values of the BH mass, larger spin values would be required to superradiate efficiently.

\section{Conclusion}
In this short note, we have shown how the recent measurements of mass and spin of the primary BH from observations of two new BBHs by LIGO/Virgo, and the observation of an X-ray binary can probe the ultralight DP in different mass ranges due to the phenomenon of superradiance. Using a combination of analytic and numerical results, we have found that the BBH observations, GW190517 and GW190426\_152155, disfavours the mass ranges $ 1.7\times 10^{-14}{\rm\ eV}<m_{A'}<7.6\times 10^{-13}{\rm\ eV}$ and $
 1.3\times 10^{-13}{\rm\ eV}<m_{A'}<4.2\times 10^{-12}{\rm\ eV}$ respectively. 
 
 It should be noted that these bounds assume only gravitational interactions. Thus, these results are valid only if the considered vector field is weakly coupled, i.e, possible non-gravitational couplings with other particles, as well as any non-trivial self-interactions should be negligible compared to the gravitational interaction. For scalar bosons, it has been shown that self-interactions or couplings of bosons to other particles can affect the superradiant instability, and significantly change the time scale required to extract a substantial amount of energy and angular momentum from the BH~\cite{PhysRevD.83.044026,Yoshino:2012kn,Yoshino:2015nsa,Rosa:2017ury,Ikeda:2018nhb,Boskovic:2018lkj,Fukuda:2019ewf,Mathur:2020aqv}. However, the effect of such interactions for massive vectors has not  been studied in detail. 
 
 There are other caveats: analysis similar to ours is plagued with our lack of understanding of formation history (thus lifetime and initial spin are largely uncertain) of a given BH binary; any given pair may not have superradiated enough if they merged too quickly or formed too close together. Moreover, the backreaction due to the DP cloud and perturbations due to the secondary BH on the Kerr geometry can affect the final results. However, this effect is expected to be not significant.~\cite{Brito_2015,East:2017ovw,Baryakhtar:2017ngi}.
 
As the sensitivity of the gravitational wave observatories improves, they will detect more or more such mergers, and thus, future observations are expected to cover larger parameter space of DP mass and possibly find the first signature of such light bosons.
% 
%%%%%%%%%%%%%%%%%%%%%%%%%%%%%%%%%%%%%%%%%%%%%%%%%%%%%%%%%%%%%%%%%%%

\begin{acknowledgments}
The authors acknowledge support through the Ramanujan Fellowship and the MATRICS grant of the Department of Science and Technology, Government of India.
\end{acknowledgments}

\bibliography{reference}
\end{document}